\documentclass[aps,prl,twocolumn,groupedaddress]{revtex4}
\usepackage{graphicx}
\usepackage{amsmath}
\usepackage{amsfonts}
\usepackage{amssymb}

\begin{document}

\title{Weak localization effects in granular metals}
\author{C. Biagini$^1$, T. Caneva$^2$, V. Tognetti$^{2,3,4}$ \& A.A. Varlamov%
$^5$ }
\affiliation{$^1$ INFM, Unit\`a di "Tor Vergata", Viale del Politecnico 1, 00133 Roma,
Italy \\
$^2$ Dip. di Fisica, Via G. Sansone 1, 50019 Sesto F.no (Firenze), Italy\\
$^3$ INFN, Sezione di Firenze, Via G. Sansone 1, 50019 Sesto F.no (Firenze),
Italy\\
$^4$ INFM, Unit\`a di Firenze, Via G. Sansone 1, 50019 Sesto F.no (Firenze),
Italy\\
$^5$ COHERENTIA-INFM, CNR, Viale del Politecnico 1, 00133, Roma, Italy}

\begin{abstract}
The weak localization correction to the conductivity of a granular
metal is calculated using the diagrammatic technique in the
reciprocal grain lattice representation. The properties of this
correction are very similar to the corresponding one in disordered
metal, with the replacement of the electron mean free path $\ell $
by the grain diameter $d$ and the dimensionless conductance $g$ by
the tunnelling dimensionless conductance $g_{T}$. In particular,
we demonstrate that at zero temperature no conducting phase can
exist for dimensions $D\leq 2$. We also analyze the WL correction
to magnetoconductivity in the weak field limit.
\end{abstract}

\date{\today }
\maketitle

Recently, the properties of granular materials have attracted special
attention \cite{BE99,ET,BELV03,AGK04}. The quantization of the electron
spectrum in small grains requires to revise the basic idea of the
quasiparticle spectrum continuity, assumed in the description of most
properties of metallic and superconducting systems. The appearance of a new
energy scale, the mean level spacing $\delta $, results in unusual
superconducting properties of such systems, possibility to observe there
specific quantum phase transitions, etc. In particular, it turns out that
the interplay between the intra-grain diffusion and inter-grain tunnelling
of electrons makes the metal-insulator transition in such \textquotedblright
quantum metal" very peculiar. In this Letter, we intend to discuss the
specifics of the weak localization corrections in granular systems.

The elastic electron relaxation rate in granular metal consists of three
contributions
\begin{equation}
\frac{1}{\tau _{el}}=\frac{1}{\tau _{imp}}+E_{T}+\Gamma ,
\label{relaxation.rate}
\end{equation}%
where $\tau _{imp}$ is the mean scattering time of electrons with
impurities, $E_{T}=v_{F}/d$ is the Thouless energy, $d$ is the
characteristic grain size, $v_{F}$ is the intra-grain Fermi velocity). The
last term is the electron inter-grain tunnelling rate: $\Gamma \sim
g_{T}\delta $, where $g_{T}=2\pi \left( t/\delta \right) ^{2}$ is the
tunnelling dimensionless conductance and $t$ is the tunnelling energy.

The character of electron motion in a metal is conveniently
classified as a function of its mean free path $\ell $\cite{note}.
The diffusion length $\ell^{(n)}_{T}=%
\sqrt{\mathcal{D}_{n}/T}$ (where $\mathcal{D}_{n}=v_{F}^{2}\tau
_{imp}/D$ is the diffusion coefficient of metal and $D$ is the
space dimensionality) separates the regions of ballistic ($\ell
\gg \ell _{T}$) and diffusive ($\ell \ll \ell _{T}$) electron
motion. When $\ell \rightarrow \hbar /p_{F}$ the metal-insulator
transition in $3D$ case takes place.
\begin{figure}[h]
\includegraphics[width=8cm]{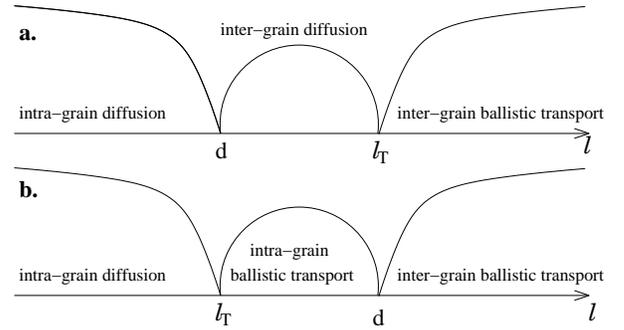}
\caption{A schematic representation of the two possible scenarios
for the electron motion in granular metals as a function of the
ratio between $d$, the grain diameter, $\ell _{T}$, the thermal
length, and $\ell $, the mean free path.} \label{l-axis}
\end{figure}
In the case of a granular metal, one can expect that the standard
WL theory, describing the precursor effects of this transition
\cite{AA84,LR85}, has to be modified in two ways. First, the
diffusion coefficient $\mathcal{D} $ here, at least in some
interval of parameters, has to be determined by the tunnelling
time $\Gamma ^{-1}$ instead of elastic scattering time $\tau
_{imp}$ and therefore we will have a different tunnelling thermal
length $\ell_T=\sqrt{{\cal D}/T}$. Second, the characteristic
grain size $d$ must appear in the theory side by side with the
diffusion length $\ell _{T}$. Two different situations are
possible. When $d\ll \ell _{T}$, the diffusive part of the $\ell
-$axis is divided to the regions of the normal intra-grain
diffusion ($\hbar
/p_{F}\ll \ell \ll d$) with the metallic diffusion coefficient \ $\mathcal{D}%
_{n}$ and of the tunnel inter-grain electron diffusion ($\ d\ll
\ell \ll \ell _{T}$) characterized by the diffusion coefficient
$\mathcal{D}=\left( \left\langle
\mathbf{x(}t\mathbf{)}^{2}\right\rangle /t\right) _{t\rightarrow
\infty }\sim \Gamma d^{2}$ (see Fig. \ref{l-axis}a). In the
opposite limit, when $d\gg \ell _{T}$, only the intra-grain
diffusion is possible, but, in its turn, the domain of ballistic
regime contains two regions: the intra-grain one with ballistic
time $\tau _{imp}$ ($\ell _{T}\ll \ell \ll d$) and the region of
inter-grain electron motion with the ballistic time $\Gamma ^{-1}$
(see Fig. \ref{l-axis}b).

We will discuss here the most interesting case of low
temperatures, $d\ll \ell _{T}$. We will also assume that the
electron motion inside a single grain is ballistic ($\ell_{bulk}
\gtrsim d$, $\ell\gg d$): this means that before tunnelling to the
neighbor grain the electron is reflected many times on the grain
boundaries ($\Gamma \ll E_{T}$). As a consequence, $\tau
_{el}\approx E_{T}^{-1}$ and the dimensionless conductance of a
single grain $g=E_{T}/\delta \gg g_{T}$. The conductance of the
entire system is given by $g^{-1}+g_{T}^{-1}\approx g_{T}^{-1}$,
what is equivalent to say that the drop of applied electric
potential occurs only inside the tunnel barrier.

We assume that the grains are almost identical, with average
diameter $d$, and form a regular lattice with lattice constant
equal to the same $d$. The coordinate of the grain center will be
labelled by the lattice variable $\mathbf{R}_{i}$. The Hamiltonian
of the system can be written as \cite{BE99}
\begin{eqnarray}
\hat{H} =\sum_{i,\mathbf{p}}\varepsilon _{\mathbf{p}}\hat{c}_{i,\mathbf{p}%
}^{\dagger }\hat{c}_{i,\mathbf{p}}+\frac{1}{2}\sum_{\left\langle i,j\right\rangle }\sum_{\mathbf{p},\mathbf{p%
}^{\prime }}\left[ t_{ij}^{\mathbf{p},\mathbf{p}^{\prime }}\hat{c}_{i,%
\mathbf{p}}^{\dagger }\hat{c}_{j,\mathbf{p}^{\prime
}}+h.c.\right]\nonumber
\end{eqnarray}%
where $\hat{c}_{i,\mathbf{p}}$ ($\hat{c}_{i,\mathbf{p}}^{\dagger
}$) is the annihilation (creation) operator of an electron in
grain $i$ with intra-grain momentum $\mathbf{p}$. The tunnelling
energy $t_{ij}^{\mathbf{p},\mathbf{p}^{\prime }}$ will be taken
equal for all bonds between nearest neighbor grains and
independent on the intra-grain momentum,
$t_{ij}^{\mathbf{p},\mathbf{p}^{\prime }}=t$. Performing the
Fourier transform with respect to such $\mathbf{R}_{i}$ \ one can
write the Hamiltonian in the representation of both intra- and
inter-grain momenta (double momentum representation):
\begin{eqnarray}
\hat{H} &=&\sum_{\mathbf{K},\mathbf{p}}\left[ \varepsilon _{\mathbf{p}}+tZ\gamma _{%
\mathbf{K}}\right] \hat{c}_{\mathbf{K},\mathbf{p}}^{\dagger }\hat{c}_{%
\mathbf{K},\mathbf{p}}  \notag \\
&+&\frac{tZ}{2}\sum_{\mathbf{K}}\sum_{\mathbf{p}\neq \mathbf{p}^{\prime
}}\gamma _{\mathbf{K}}\left[ \hat{c}_{\mathbf{K},\mathbf{p}}^{\dagger }\hat{c%
}_{\mathbf{K},\mathbf{p}^{\prime }}+h.c.\right] .  \label{2}
\end{eqnarray}%
Here $\mathbf{K}$ is the quasi-momentum belonging to the reciprocal to $%
\mathbf{R}_{i}$ lattice and it varies in the first Brillouin zone. The
lattice structure factor $\gamma _{\mathbf{K}}=Z^{-1}\sum_{\mathbf{\mu =1}%
}^{Z}e^{i\mathbf{K}\cdot \mathbf{d}_{\mu }}$; $\mathbf{d}_{\mu }$
are the vectors connecting the center of selected grain with the
nearest neighbor sites, $Z$ is the coordination number. For a
simple cubic lattice, the vectors $\mathbf{d}_{\mu }$ have one
component equal to $\pm d$ and all the others equal to zero. For
sake of simplicity, we will restrict to this case; the extension
to generic lattices is straightforward. In simple cubic lattices,
$\gamma _{%
\mathbf{K}}=\frac{1}{D}\sum_{\alpha =1}^{D}\cos \left(
K_{\alpha}d\right)$. From Eq. (\ref{2}) we can define the single
electron Green's function in the double momentum representation
as:
\begin{eqnarray}
G_{\mathbf{K}}\left( \mathbf{p},\varepsilon _{n}\right)  &=&\frac{1}{i\tilde{%
\varepsilon}_{n}-\xi _{\mathbf{p}}-Zt\left( 1-\gamma _{\mathbf{K}}\right) },
\label{3}
\end{eqnarray}%
with $\tilde{\varepsilon}_{n}=\varepsilon _{n}+(2\tau _{el})^{-1}\mathrm{sign%
}\varepsilon _{n}$ and $\varepsilon _{n}=\pi T\left( 2n+1\right) $ as
fermionic Matsubara's frequency.

Recalling that the electric field is negligible inside the grains and
differs from zero only inside the barriers, in presence of the vector
potential $\mathbf{A}$ one can write the $\alpha -$th component of the
electrical current operator in the imaginary time $\tau $ as
\begin{eqnarray}
\hat{J}_{\alpha }\left( \tau \right)  &=&i\frac{etd}2\sum_{\mathbf{K},%
\mathbf{p},\mathbf{p}^{\prime }}\left[ e^{iK_{\alpha }d}\hat{c}_{%
\mathbf{K},\mathbf{p}}^{\dagger }\left( \tau \right) \hat{c}_{\mathbf{K},%
\mathbf{p}^{\prime }}\left( \tau \right) -h.c.\right]   \notag \\
&-&\frac{e^{2}d}2\left( \mathbf{A}\left( \tau \right) \cdot \mathbf{d}%
\right) \hat{H}_{T}\left( \tau \right) .  \label{current}
\end{eqnarray}%
The linear response function, expressed as the second derivative
of the partition function, is given by
\begin{eqnarray}
\mathcal{K}_{\alpha ,\alpha ^{\prime }}\left( \tau \right)  &=&-\frac{1}{%
\mathcal{Z}[0]}\left. \frac{\delta ^{2}\mathcal{Z}[A]}{\delta A_{\alpha
}\left( \tau \right) \delta A_{\alpha ^{\prime }}\left( 0\right) }%
\right\vert _{\mathbf{A}\rightarrow 0}=  \notag \\
&=&\Pi _{\alpha \alpha ^{\prime }}\left( \tau \right)
-e^{2}d^{2}\delta _{\alpha \alpha ^{\prime }}\delta \left( \tau
\right) \left\langle \hat{H}_{T}\right\rangle _{0},
\label{response.function}
\end{eqnarray}%
where the current-current correlation function is expressed via the current
operator as
\begin{equation*}
\Pi _{\alpha ,\alpha ^{\prime }}\left( \tau \right) =-\left\langle \hat{T}%
_{\tau }\hat{J}_{\alpha }\left( \tau \right) \hat{J}_{\alpha ^{\prime
}}\left( 0\right) \right\rangle _{0}.
\end{equation*}%
The thermal average $\left\langle ...\right\rangle _{0}$ shall be
performed with the diagonal Hamiltonian, the first line of Eq.
(\ref{2}). We are interested in the diagonal components of
conductivity tensor:
\begin{eqnarray}
\Pi _{\alpha ,\alpha }\left( \omega _{\nu }\right)
&=&2e^{2}d^{2}\left\vert t\right\vert ^{2}\sum_{\mathbf{K}}\sin
^{2}\left( K_{\alpha
}d\right)   \notag \\
&\times &T\sum_{\varepsilon _{n}}\sum_{\mathbf{p},\mathbf{p}^{\prime }}G_{%
\mathbf{K}}\left( \mathbf{p},\varepsilon _{n+\nu }\right) G_{\mathbf{K}%
}\left( \mathbf{p}^{\prime },\varepsilon _{n}\right) .  \label{longitudinal}
\end{eqnarray}%
where $\varepsilon_{n+\nu}=\varepsilon_n+\omega _{\nu }$ and
$\omega _{\nu }=2\pi T\nu $ is the bosonic Matsubara's frequency.
We can formulate the following rules of diagrammatic technique in
the double
momentum representation: $1.$ at each external vertex attach a factor $\hat{v%
}_{\alpha }=etd\sin \left( K_{\alpha }d\right) $; $2.$
at each straight line attach a single electron Green's function $G_{\mathbf{K%
}}\left( \mathbf{p},\varepsilon _{n}\right) $; $3.$ sum over all internal
momenta and Matsubara's frequencies; $\ 4.$ impose energy and lattice
momentum conservation at each vertex.
\begin{figure}[tbp]
\includegraphics[width=8cm]{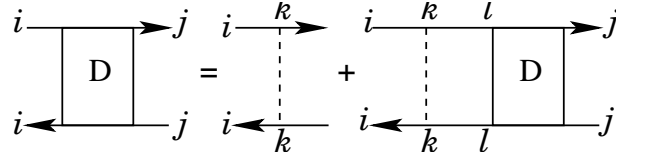}
\caption{Dyson equation for the diffusion vertex part in a
granular metal in the direct space. The thick lines are the double
Fourier transform of the fermionic Green's functions Eq.
(\protect\ref{3}), the dashed lines represent single impurity
scattering and the initial and final grains are reported as roman
indices $i$, $j$ etc. The Cooperon is obtained from the diffuson
via time-reversal transformation on a fermionic Green's function
\protect\cite{AA84}.} \label{Dyson}
\end{figure}
Now the impurity averaging of Eq. (\ref{longitudinal}) can be
performed. The Cooperon vertex \cite{AA84} corresponding \ to the
granular metal can be obtained from the Dyson's equation reported
in Fig. (\ref{Dyson}). Other possibility is to renormalize the
standard intra-grain Cooperon by means of introduction in the
corresponding diagrams of a self-energy correction appearing due
to tunnelling, as it is done in Ref. \cite{BE99}. Both approaches
in the assumptions made above turn out to be completely equivalent
and lead to the expression:
\begin{equation*}
\mathcal{C}_{\mathbf{Q}}\left( \omega _{\nu }\right) =\frac{1}{2\pi \tau
_{el}^{2}\nu _{F}}\frac{1}{\left\vert \omega _{\nu }\right\vert +2\Gamma
\left( 1-\gamma _{\mathbf{Q}}\right) }.
\end{equation*}%
In the latter expression appears the exact value for the tunnelling rate: $%
\Gamma =Zg_{T}\delta $. In the expression $\mathcal{D}=\left[ 2\Gamma \left(
1-\gamma _{\mathbf{Q}}\right) /\left\vert \mathbf{Q}\right\vert ^{2}\right]
_{\mathbf{Q}\rightarrow 0}=\Gamma d^{2}$ one can also recognize the
effective \textquotedblright tunnelling diffusion constant".
\begin{figure}[tbp]
\includegraphics[width=6cm]{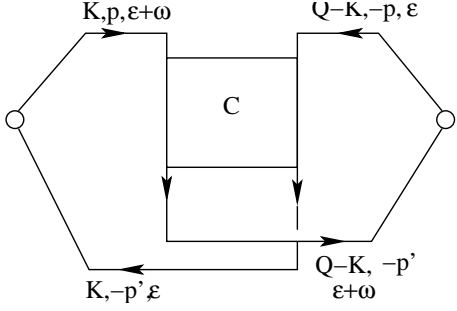}
\caption{WL correction to the conductivity in the double momentum
representation. The solid lines are single electron temperature Green's
functions $G_{\mathbf{K}}\left( \mathbf{p},\protect\varepsilon _{n}\right) $%
; the external velocity vertices are $\hat{v}_{\protect\alpha
}=etd\sin \left( K_{\protect\alpha }d\right) $; the
shaded box is the Cooperon $\mathcal{C}_{\mathbf{Q}}\left( \protect\omega _{%
\protect\nu }\right) $.}
\label{Cooperon}
\end{figure}

Now one can directly calculate the diagram reported in Fig. (\ref{Cooperon}%
). In our assumptions, see Eq. (\ref{relaxation.rate}), the
Green's function can be considered independent on $\mathbf{K}$
when integrating over $\mathbf{p}$, because its behavior is
completely determined by the pole due to the impurity (or grain
boundaries) scattering, related to $\tau _{el}^{-1}$, just as in
the case of the diffusive limit for a bulk system. Performing the
frequency summation, the $\mathbf{p}$ and $\mathbf{p}^{\prime }$
integration and the sum over the lattice momentum ${\bf K}$ one
finds
\begin{eqnarray*}
\frac{\mathcal{K}_{\alpha \alpha }^{WL}\left( \omega _{\nu
}\right) }{\omega _{\nu }}=-\frac{e^{2}d^{2}}{4\pi\nu
_{F}}g_{T}\sum_{\mathbf{Q}}\frac{\cos \left( Q_{\alpha }d\right)
}{\left\vert \omega _{\nu }\right\vert +2\Gamma \left( 1-\gamma
_{\mathbf{Q}}\right) }
\end{eqnarray*}%
At this point, we can find the WL correction to the conductivity as
\begin{equation}
\frac{\delta \sigma _{(D)}^{WL}}{\sigma _{(D)}^{n}}=-\frac{1}{\pi Zg_{T}}\sum_{%
\mathbf{Q}}\frac{\cos \left( Q_{\alpha }d\right) }{1-\gamma _{%
\mathbf{Q}}}.  \label{weak.loc.final}
\end{equation}

In the case of a bulk granular system ($D=3$) the sum converges
and the correction is finite. The metal-insulator phase transition
can be observed at a critical value $Zg_{T}^{cr}\sim{\cal O}(1)$.

In the case of granular film or wire ($D\leq 2$) the WL correction Eq. (\ref%
{weak.loc.final}) diverges at small $\mathbf{Q\rightarrow }0$. This fact
indicates that at zero temperature these systems cannot exhibit metallic
properties for any value of the dimensionless conductance $g_{T}$. At finite
temperatures, a natural cut-off of Eq. (\ref{weak.loc.final}) is provided by
the phase-breaking rate $\gamma _{\varphi }=\left( \Gamma \tau _{\varphi
}\right) ^{-1}=\left( d/L_{\varphi }\right) ^{2}$ \cite{AA84}. It is related
to the phase-coherence length $L_{\varphi }=\sqrt{\mathcal{D}\tau _{\varphi }%
}$, which tends to infinity when $T\rightarrow 0$. In result
\begin{eqnarray}
\frac{\delta \sigma _{(1)}^{WL}}{\sigma _{(1)}^{n}} \approx
-\frac{2}{Z\pi ^{3}g_{T}}\frac{1}{\sqrt{\gamma _{\varphi }}};
\frac{\delta \sigma _{(2)}^{WL}}{\sigma _{(2)}^{n}} \approx
-\frac{1}{Z\pi^2 g_{T}}\ln \frac{\pi ^{2}}{\gamma _{\varphi }}.
\label{weak.loc.2d}
\end{eqnarray}%
Eqs. (\ref{weak.loc.2d}) permit to define the localization length
$\xi _{(D)}^{loc}$ at which the correction becomes of the order of
$\sigma _{(D)}^{n}$:
\begin{equation*}
\frac{\xi _{(1)}^{loc}}{d}\approx
\frac{Z\pi^{3}}{2}g_{T};~~~\frac{\xi _{(2)}^{loc}}{d}\approx
e^{\frac{Z\pi^2 }{2}g_{T}}/\pi .
\end{equation*}%
It is worth to stress that Eq. (\ref{weak.loc.final}) does not contain $\tau
_{el}^{-1}$ but only lattice parameters as $g_{T}$ and the coordination
number $Z$: the intra-grain dynamics, as expected, simply drops out of the
calculation. The conductivity and its corrections are related to the
diffusion on the grain lattice, and the mechanism of the momentum
randomization between different grains is not crucial: the electron dynamics
at low temperature can be thought as that of a random walker on a lattice.
This picture is fully consistent with the existing WL theory \cite%
{AA84,LR85}, and it is in agreement with previous experimental
findings in granular metals \cite{exp}.

More intriguing are the properties of the WL correction to the
magnetoresistance of granular metal. It is well known that quantum
corrections to conductivity in disordered metal are very sensitive
to the magnetic field: in fact, its presence disturbs the phase
coherence of electrons moving along the self-intersecting
trajectories, suppressing the WL correction and leading to the
appearance of the anomalous negative magnetoresistance
\cite{AA84}. In the following, we will show how such a correction
manifests itself in the case of granular metal.

To calculate the WL contribution to magnetoresistance, it is necessary to
rewrite Eq. (\ref{weak.loc.final}) in the direct space:
\begin{equation*}
\frac{\delta \sigma _{(D)}^{WL}}{\sigma _{(D)}^{n}}=-\frac{2}{Zg_{T}}\Gamma
\frac{\tilde{C}_{i,i+\alpha }+\tilde{C}_{i+\alpha ,i}}{2}
\end{equation*}%
which is independent on $i$, depending only on the inter-grain spacing $d$. $%
\alpha $ represents the bond along the direction of the current.
Here $\tilde{C}_{\mathbf{Q}}^{-1}=\left( 2\pi \tau _{el}^{2}\nu
_{F}C_{\mathbf{Q}}\right) ^{-1}=-i\omega +2\Gamma \left( 1-\gamma _{\mathbf{Q%
}}\right) $. Let us notice that this form underlines the fact that
transport is due only to the potential drop inside the barrier
separating two grains $i $ and $i+\alpha $. In the presence of a
magnetic field, the Cooperon wavefunction is given by the solution
of the equation
\begin{equation}
\left(4\Gamma\right)\left(1-\gamma_{{\bf Q}+2e{\bf
A}}\right)\psi_i\left({\bf r}\right)=E\psi_i\left({\bf r}\right);
\end{equation}
Moreover, also the intra-grain Cooperon will be renormalized by
the presence of the magnetic field, acquiring a mass term equal
to\cite{BE99} ${\cal
E}_0\left(H\right)=\frac25\left(\frac{\pi\phi}{\phi_0}\right)^2E_T$,
where $\phi=Hd^2$ is the magnetic flux threaded through a single
spherical grain and $\phi_0=\pi/e$ is the flux quantum. When the
field satisfies the inequality $d\ll \ell _{H}=\left(
eH\right)^{-1/2}$, or $\phi\ll\phi_0$, we have
\begin{eqnarray}
&&\tilde{C}_{ij }\left( \mathbf{r},\mathbf{r}^{\prime },\omega
\right)\\
&=&\sum_{Q_{\parallel},Q_\perp,n}\frac{\psi _{i,Q_{\parallel }Q_\perp n}\left( \mathbf{%
r}\right) \psi _{j,Q_{\parallel}Q_\perp n}^{\star }\left( \mathbf{r}%
^{\prime }\right) }{-i\omega +\Omega _{c}\left( n+\frac{1}{2}\right) +%
\mathcal{D}Q_{\parallel }^{2}+\frac{1}{\tau _{\varphi }}+{\cal
E}_0\left(H\right)}\nonumber \label{Cooperon.H}
\end{eqnarray}%
where $\Omega _{c}=4\mathcal{D}/\ell _{H}^{2}$ is the Cooperon
cyclotron energy. $Q_{\parallel }$ is the momentum along the
magnetic field and $\left(n,Q_{\perp}\right)$ are the quantum
numbers of the Landau basis.

The most interesting case is the two-dimensional geometry, with
the magnetic field applied across the plane of the
sample\cite{AA84} for which the WL correction is:
\begin{equation*}
\frac{\delta \sigma _{(2)}^{WL}}{\sigma _{(2)}^{n}}=-\frac{\cos \left( \frac{%
\Omega _{c}}{2\Gamma }\right) }{\pi Zg_{T}}\sum_{n=0}^{n_{\max }}\frac{1}{n+%
\frac{1}{2}+\left(\gamma _{\varphi }+\frac{{\cal
E}_0}\Gamma\right)\frac{\Gamma }{\Omega _{c}}};
\end{equation*}%
Since the sum over Landau levels is evidently divergent at the
upper limit we introduced the cut-off parameter $n_{\max }=\pi
^{2}\Gamma /\Omega _{c}\gg 1$, at which the cyclotron frequency
becomes of the order of the zone edge, $n_{\rm max}\Omega _{c}\sim
{\cal D}\left(\pi/d\right)^2$. The final result reads
\begin{equation*}
\frac{\delta \sigma _{(2)}^{WL}}{\sigma _{(2)}^{n}}=-\frac{1}{\pi Zg_{T}}%
\mathcal{F}\left( \phi,\gamma _{\varphi }\right) ,
\end{equation*}%
where we expressed the magnetic field in terms of the magnetic
flux,
\begin{eqnarray}
\mathcal{F}\left( \phi,\gamma _{\varphi }\right)  &=&\cos \left(
2\pi\frac{\phi}{\phi_0}\right) \Bigg[\psi \left( \frac{\pi}4\frac{\phi_0}{\phi}+%
\frac{1}{2}+\frac\pi{10}\frac{\phi}{\phi_0}\frac{E_T}{\Gamma}\right)   \notag \\
&-&\psi \left( \frac{\gamma _{\varphi }}{4\pi}\frac{\phi_0}{\phi}+\frac{1}{2}
+\frac\pi{10}\frac{\phi}{\phi_0}\frac{E_T}{\Gamma}%
\right) \Bigg].  \label{magnetoresistance.2d.final.1}
\end{eqnarray}%
where $\psi (x)$ is the digamma function. The WL correction to the
magnetoresistance $\delta \sigma \left( H\right)=\delta \sigma
^{WL}\left( H\right) -\delta \sigma ^{WL}\left( 0\right)$ is
obtained as
\begin{eqnarray}
\delta \sigma \left( H\right)=-\frac{\sigma _{0}}{\pi
Zg_{T}}\left[ \mathcal{F}\left( H,\gamma _{\varphi }\right) -\ln
\frac{\pi ^{2}}{\gamma _{\varphi }}\right] . \label{delta.sigma.H}
\end{eqnarray}%
One more energy scale shows up in Eq. (\ref{delta.sigma.H}) with
respect to the bulk case, namely the Thouless energy. In the limit
of very weak fields, $\phi/\phi_0\ll\sqrt{1/4E_T\tau_{\varphi}}$,
this energy scale is not observable in the magnetoresistance: the
leading singular correction reduces to
\begin{eqnarray*}
\frac{\delta \sigma \left( H\right) }{\sigma _{0}}&\approx&\frac{2\pi}{3}\frac{1%
}{Zg_{T}}\left( \frac{\phi}{\phi_0}\right) ^{2}\propto H^2
\end{eqnarray*}%
which corresponds to the anomalous magnetoresistance of the
standard theory.  The granular behavior deviates from the bulk one
at fields such that $\phi/\phi_0\sim \Gamma/E_T$ where the
intra-grain term starts to dominate in the second digamma function
in $\mathcal{F}\left( H,\gamma _{\varphi }\right)$: for
$\sqrt{1/4E_T\tau_{\varphi}}\ll\phi/\phi_0\ll\gamma_{\varphi}$,
the magnetoresistance correction acquires the logarithmic form
\begin{eqnarray*}
\frac{\delta \sigma \left( H\right) }{\sigma _{0}} =\frac2{\pi
Zg_T}\log\left(\sqrt{\frac{E_T}{\Gamma}}\frac{\phi}{\phi_0}\right).
\end{eqnarray*}
Larger fields are out of the range of our approximated approach,
in which the intra-grain Landau levels, with the spectrum
$\omega_c\left(n+1/2\right)$ with $\omega_c=4{\cal D}_n/\ell_H^2$,
start to significantly contribute to the Cooperon wavefunction.
Let us notice finally that the ultraviolet cut-off $n_{\rm
max}\sim E_T/\omega_c=\Gamma/\Omega_c$ remains the same.

In summary, we developed a diagrammatic technique in a
double-momentum representation for transport in granular metals.
Using this technique, the weak localization corrections to the
conductivity arise in a natural way and an explicit calculation
shows the same low-temperature behavior as in bulk metals, but
with the diffusion constant ${\cal D}_n$ replaced by the effective
tunnelling diffusion constant ${\cal D}=\Gamma d^2$ and the mean
free path $\ell$ by the average grain diameter $d$. Our result
agrees with Eq. (13) of Ref. \cite{BLV04} in the ${\bf
Q}\rightarrow0$ limit; however, our technique underlines the
presence of the grain lattice, represented by the cosine factor in
Eq. (\ref{weak.loc.final}), reminiscent of the lattice structure
factor $\gamma_{\bf Q}$. We also give an estimate of the
magnetoresistance correction for very weak fields.

\acknowledgements We deeply thank illuminating discussions with A.
Cuccoli, Rosario Fazio, R. Ferone, I.V. Lerner, I.V. Yurkevich and
B.L. Al'tshuler. C.B. acknowledges the hospitality of the
University of Birmingham, where this work has started.


\begin{thebibliography}{9}
\bibitem{note} In a granular metal, $\ell$ coincides with the bulk mean
free path $\ell_{bulk}$ up to the grain size $d$. If
$\ell_{bulk}\gtrsim d$, the grain is in the ballistic limit. On
the other hand, the electron will loose memory of its initial
momentum state after a time of the order of $\tau_{el}$: Eq.
(\ref{relaxation.rate}) with the assumption
$\Gamma\ll\max\{1/\tau_{imp},E_T\}$ implies that the electron
relaxes its momentum only due to intra-grain effects. We will
define as a new mean free path just the new length scale, that
appears only in finite systems, over which the electron looses
completely the memory of its initial momentum state, $\ell\equiv
v_F\tau_{el}$.
\end{thebibliography}
\end{document}